\documentclass[prb,amsmath,amssymb,twocolumn,showpacs,floatfix]{revtex4}
\usepackage{graphicx}
\usepackage{dcolumn}
\usepackage{bm}

\begin{document}

\title{Magnetic properties of the helimagnet Cr$_{1/3}$NbS$_2$ observed by $\mu$SR}

\author{D.~Braam,$^1$ C.~Gomez,$^1$ S.~Tezok,$^1$ E.V.L.~de Mello,$^2$ L. Li,$^3$ D. Mandrus,$^{3,4}$ Hae-Young Kee,$^{5,6}$ and J.E.~Sonier,$^{1,6}$}

\affiliation{$^1$Department of Physics, Simon Fraser University, Burnaby, British Columbia V5A 1S6, Canada \\
$^2$Instituto de F\'{i}sica, Universidade Federal Fluminense, Niter\'{o}i, RJ 24210-340, Brazil \\
$^3$Department of Materials Science and Engineering, University of Tennessee, Knoxville, Tennessee 37996, USA \\ 
$^4$Oak Ridge National Laboratory, Oak Ridge, Tennessee 37831, USA \\
$^5$Department of Physics, University of Toronto, Ontario M5S 1A7, Canada \\
$^6$Canadian Institute for Advanced Research, Toronto, Ontario M5G 1Z8, Canada}

\date{\today}

\begin{abstract}
We have performed muon spin rotation/relaxation ($\mu$SR) measurements on single crystals of the chiral helimagnet 
Cr$_{1/3}$NbS$_2$ at zero to low magnetic field. The transition from the paramagnetic to helical magnetically ordered
phase at zero field is marked by the onset of a coherent oscillation of the zero-field muon spin polarization below a
critical temperature $T_c$. An enhancement of the muon spin precession frequency is observed below $T \! \sim \! 50$~K,
where anomalous behavior has been observed in bulk transport measurements. The enhanced precession frequency
indicates a low-temperature modification of the helical magnetic structure. 
A Landau free energy analysis suggests that the low-temperature change in the magnetic structure is caused
by a structural change. We also suggest a longer periodicity of helicity below $T \! \sim \! 50$~K, which can be verified 
by neutron scattering experiments.    
\end{abstract}

\maketitle
Cr$_{1/3}$NbS$_2$ is an itinerant chiral helimagnet (CHM) with an hexagonal non-centrosymmetric crystal structure.\cite{Miyadai:83} 
In crystals belonging to a non-centrosymmetric space group, CHM order is generally understood to be a consequence of competing 
symmetric Heisenberg exchange and Dzyaloshinsky-Moriya (DM) interactions. The CHM order in Cr$_{1/3}$NbS$_2$ continuously evolves 
into an incommensurate chiral magnetic solition lattice (CSL) upon application of a magnetic field perpendicular to the helical $c$-axis 
(${\bf H} \! \perp \! \hat{c}$).\cite{Togawa:12,Togawa:13}
The CSL consists of periodic domains of ferromagnetically ordered spins in the $ab$ plane separated by 360$^\circ$ spin domain
walls. The period of the CSL increases monotonically with the magnitude of the applied field, resulting in a continuous
phase transition to a commensurate ferromagnetic (FM) state. The ability to easily tune the size of the magnetic domains of the 
CSL, and hence the magnetic potential experienced by the spins of the intinerant electrons, makes Cr$_{1/3}$NbS$_2$ an appealing
candidate for spintronics applications.

At elevated temperatures there is a transition to a paramagnetic (PM) state. Reported values of the PM-to-CHM phase transition 
temperature $T_c$ vary between 118~K (Ref.~\onlinecite{GhimireThesis}) and 133.5~K (Ref.~\onlinecite{BornsteinThesis}).
While the long-range magnetically ordered phases of Cr$_{1/3}$NbS$_2$ have been established, the details of 
the transition from the PM phase to the CHM or CSL phases are unresolved. 
Anomalies in the magnetic field dependence of the resistivity, and the 
temperature dependences of the Seebeck and Hall coefficients near $T_c$ have been reported.\cite{Ghimire:13,Parkin:80}
In particular, at $T_c$ a signficant negative magnetoresistance (with ${\bf H} \! \perp \! \hat{c}$) has been observed,
with no saturation of the resistivity at fields where both the resistivity and magnetization saturate at
lower temperatures.\cite{Ghimire:13} The Seebeck and Hall coefficients also show a pronounced minimum near $T_c$, indicative of changes to the electronic 
structure and/or the effects of the magnetism on the mobility of the charge carriers.  
Likewise, anomalous behavior has been observed in magnetoresistance, thermopower and Hall resistivity data below 
$T \! \sim \! 50$~K.\cite{BornsteinThesis} At these low temperatures the magnetoresistance at high fields becomes
anisotropic, there is a rapid decrease of the Seebeck coefficient with decreasing temperature, and the field dependence
of the Hall resistivity drastically changes. To gain futher insight into the sources of the transport anomalies, we have used 
the $\mu$SR technique to investigate the nature of the accompanying magnetism. 
       
Our $\mu$SR measurements were performed at TRIUMF on a $\sim \! 50$~mm$^2$ mosaic consisting of seven $\sim \! 0.25$~mm thick 
plate-like Cr$_{1/3}$NbS$_2$ single crystals, which were grown as described in Ref.~\onlinecite{Ghimire:13}. The positive muon ($\mu^+$) 
beam was directed at the large face of the mosaic, with the linear momentum of the beam parallel to the $c$-axis of the crystals 
($z$-direction). The initial muon spin polarization {\bf P}$(t \! = \! 0)$ for all measurements was rotated to be perpendicular to 
the $c$-axis ($x$-direction). Data was collected with zero applied magnetic field (ZF), and for a transverse field (TF) [{\it i.e.} 
perpendicular to {\bf P}$(t \! = \! 0)$] applied perpendicular to the $c$-axis ($y$-direction). Note that for the desired 
${\bf H} \! \perp \! \hat{c}$ orientation the field must be applied perpendicular to the beam, since the crystals are too thin
to orient with the beam shining on the $a$-$c$ or $b$-$c$ plane faces. The ensuing beam deflection by
the magnetic force exerted on the incoming $\mu^+$ limited the applied field strength to $H \! \leq \! 600$~Oe.
\begin{figure}
\centering
\includegraphics[width=9.0cm]{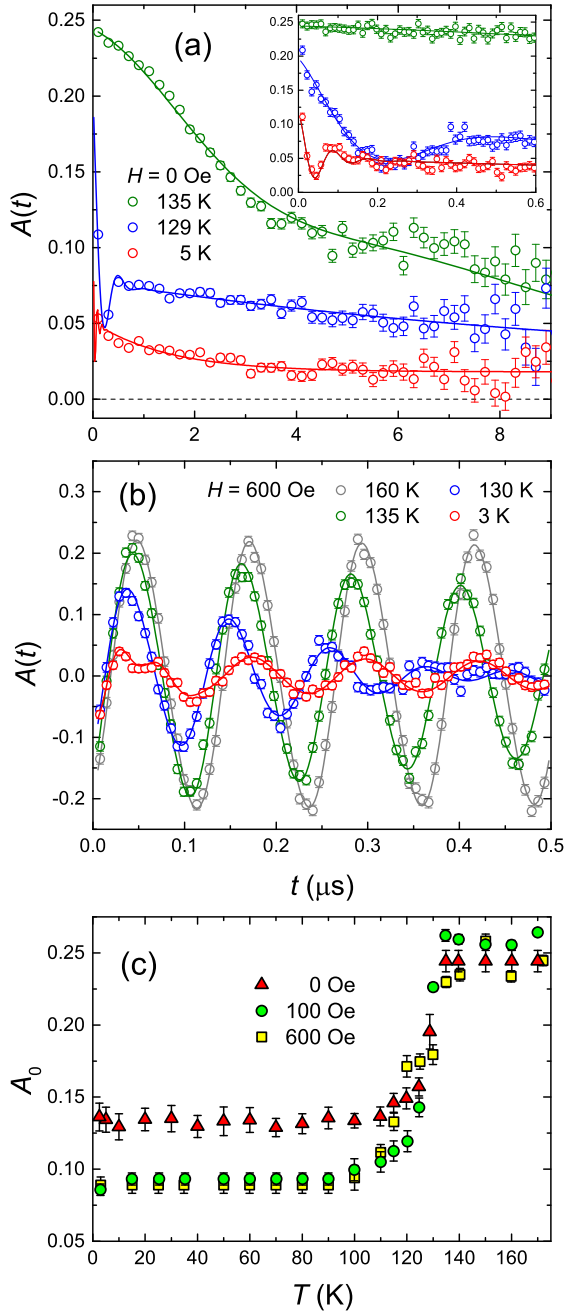}
\caption{(Color online) (a) Representative ZF-$\mu$SR asymmetry spectra of Cr$_{1/3}$NbS$_2$ recorded
at different temperatures. The inset shows the asymmetry spectra at early times, with the data packed
into smaller time bins. The solid curves through the data points are fits to 
Eqs.~(\ref{eq:PM}) or (\ref{eq:CHM}). (b) Representative TF-$\mu$SR asymmetry spectra of Cr$_{1/3}$NbS$_2$ recorded
at different temperatures with $H \! = \! 600$~Oe. The solid curves through the data points are fits to 
Eq.~(\ref{TF}). (c) Temperature dependence of the total initial asymmetry $A_0$.}
\label{fig1}
\end{figure}  

Figure~\ref{fig1}(a) shows representative ZF-$\mu$SR asymmetry spectra.
The data for $135 \! \leq \! T \! \leq \! 170$~K were best fit to
\begin{equation}
A(t) = \sum_{i = 1}^{2} A_i G_{\rm KT}(\Delta_i,t)e^{-\lambda t} + A_{\rm b} \, ,
\label{eq:PM}
\end{equation}
where $G_{\rm KT}(\Delta_i,t)$ is a static Gaussian Kubo-Toyabe relaxation function, and $A_{\rm b}$
is a small time-independent background signal. The fits yield $A_1/A_2 \! \approx \! 2$, 
$\Delta_1 \! = \! 0.091(4)$~$\mu$s$^{-1}$, $\Delta_2 \! = \! 0.414(5)$~$\mu$s$^{-1}$,
and a relaxation rate $\lambda$ that exhibits a divergent increase as the temperature is lowered. Based on these
results and nuclear dipole field calculations, we attribute the two sample components to
a muon stopping site near a S atom and a site near Nb. The amplitude ratio $A_1/A_2$ implies
that the probability of the muon stopping at the S site is approximately twice that of the Nb site.  

\begin{figure}
\centering
\includegraphics[width=8.0cm]{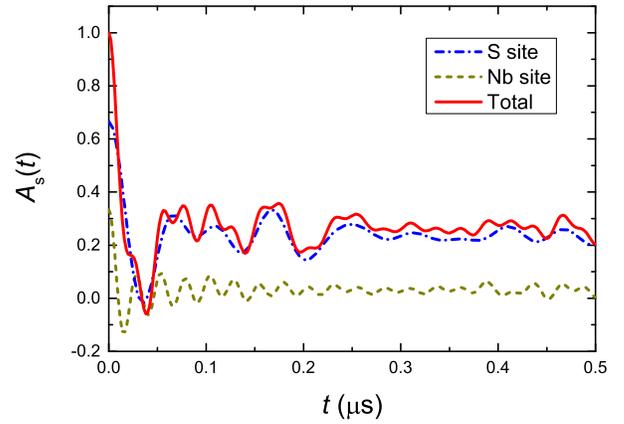}
\caption{(Color online) Simulations of the sample (s) contribution to the ZF-$\mu$SR signal for Cr$_{1/3}$NbS$_2$ 
in the CHM ground state. The dashed curves show the contribution from muons stopping near Nb at (0.77, 0.32, 0.96) and 
near S at (0.37, 0.78, 0.6), where the coordinate values are multiples of the respective lattice constants 
$a \! = \! b \! = \! 5.7 $~\AA, and $c \! = \! 12.1$~\AA. The Cr magnetic moment is assumed to be 4.4~$\mu_B$.
The occupation probability of the muon site near S is taken to be twice that of the site near Nb.}
\label{fig2}
\end{figure}

\begin{figure}
\centering
\includegraphics[width=9.0cm]{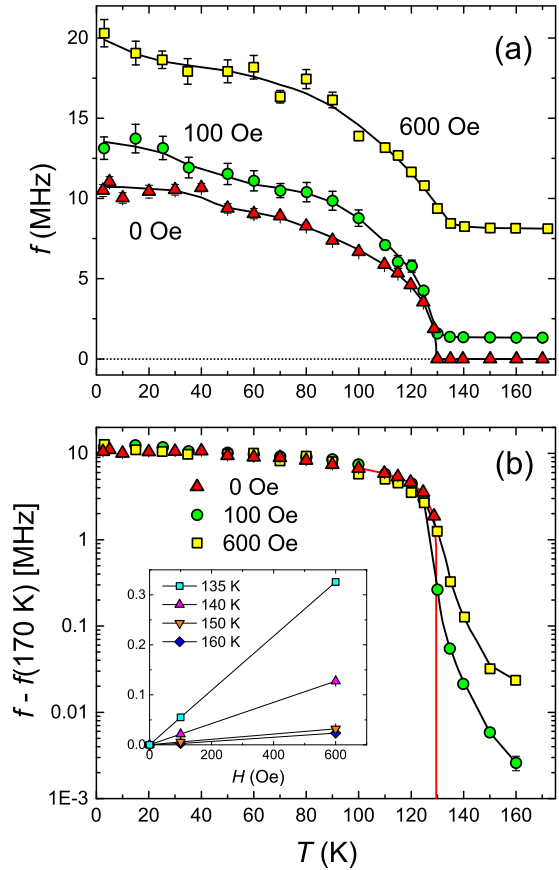}
\caption{(Color online) Temperature dependence of (a) the muon spin precession frequency $f$ observed for
different applied magnetic field strengths, and of (b) the frequency shift $f(T) \! - \! f(T \! = \! 170$~K) plotted
with a logarithmic vertical scale. The black solid curves through the data points in both panels are guides to the eye. 
The red curve in (b), which extends off the logarithmic scale, comes from a fit of the high-temperature data 
for $H \! = \! 0$~Oe to $f(T) \! = \! f(0)(1-T/T_c)^n$, yielding $T_c \! = \! 129.58(4)$~K and $n \! = \! 0.366(6)$.
The inset shows the magnetic field dependence of $f(T) \! - \! f(T \! = \! 170$~K) at temperatures above $T_c$.}
\label{fig3}
\end{figure}

For $T \! \leq \! 130$~K the relaxation rate 
is sufficiently fast that some of the measured ZF-$\mu$SR signal is damped out in the initial dead time of 
the $\mu$SR spectromter, resulting in a reduction of the initial amplitude $A_0 \! = \! A_1 + A_2 + A_{\rm b}$ [see Fig.~\ref{fig1}(c)]. 
The asymmetry spectra at these temperatures were fit to
\begin{equation}
A(t) = A_{1} e^{-\Lambda_1 t} \cos(2 \pi f t)  + A_{2} e^{-\Lambda_2 t} + A_{\rm b} \, .
\label{eq:CHM}
\end{equation}
The oscillating term is indicative of magnetic order, and the spin precession frequency $f$ is related
to the magnitude of the mean field $B_{\mu}$ sensed by the muon ensemble, where $f \! = \! (\gamma_{\mu}/2 \pi) B_{\mu}$
and $\gamma_{\mu}/2 \pi \! = \! 135.54$~MHz/T is the muon gyromagnetic ratio. Despite there being two muon stopping
sites, we could not resolve two distinct frequencies. This is understandable from simulations of the contribution
of each of these sites to the ZF-$\mu$SR signal (see Fig.~\ref{fig2}), where the signal from muons stopping near
Nb is smaller and rapidly damped. The temperature dependence of $f$ indicates 
a magnetic transition at $T_c \! = \! 129.6$~K (see Fig.~\ref{fig3}), which 
is close to previous estimates of the Curie temperature of Cr$_{1/3}$NbS$_2$.\cite{Miyadai:83}

Figure~\ref{fig1}(b) shows representative TF-$\mu$SR spectra for ${\bf H} \! \perp \! \hat{c}$ and $H \! = \! 600$~Oe.
The external field adds vectorially to the local field generated by the spin structure. At all 
temperatures the TF-$\mu$SR spectrum is well described by the sum of sample and background contribution of the following form
\begin{equation}
A(t) = A_{\rm s} e^{-\Lambda t} \cos(2 \pi f t) + A_{\rm b}e^{-\sigma^2 t^2} \cos(2 \pi f_{\rm b} t) \, ,
\label{TF}
\end{equation}
where $A_0 \! = \! A_{\rm s} + A_{\rm b}$ is the total signal amplitude. There is a rapid signal depolarization 
below $T_c$, which manifests itself in a reduction of $A_0$ that is more severe than with $H \! = \! 0$ [see Fig.~\ref{fig1}(c)]. 
Bulk magnetization
measurements suggest that the CSL state is not fully formed until $H \! \approx \! 850$~Oe.\cite{Ghimire:13}
A distortion of the CHM order into an imhomogeneous helicoid is consistent with the enhanced depolarization observed
here at weaker fields.  
As shown in Fig.~\ref{fig3}(a), the temperature dependence of the spin precession frequency 
$f$ is similar to that observed with $H \! = \! 0$, but shifted in value by the applied field.
There is some enhancement of $f$ at low temperatures, which is indicative of a structural change and/or modification 
of the spin structure. At $H \! = \! 0$~Oe and $H \! = \! 100$~Oe this occurs below $T \! \approx \! 50$~K, where as 
mentioned earlier, anomalous behavior has been observed in transport measurements.     
 
Figure~\ref{fig3}(b) shows the temperature dependence of the frequency difference $f(T) \! - \! f(T \! = \! 170$~K). 
With $H \! > \! 0$, the spin precession frequency increases as the zero-field magnetic transition temperature $T_c$ is 
approached from above. As shown in the inset of Fig.~\ref{fig3}(b), $f(T) \! - \! f(T \! = \! 170$~K$) \! \propto \! H$
at temperatures $T_c \! < \! T \! < \! 170$~K, where the proportionality constant is roughly proportional to $1/(T-T_c)$.
Hence the observed frequency shift above $T_c$ is consistent with the low-field response of a paramagnet,
whereby the induced magnetiztion is linearly dependent on $H$. A full and temperature-independent
amplitude $A_0$ is observed above $T_c$ [see Fig.~\ref{fig1}(b)], indicating that the paramagnetism occurs
throughout the sample volume.

\begin{figure}
\centering
\includegraphics[width=8.0cm]{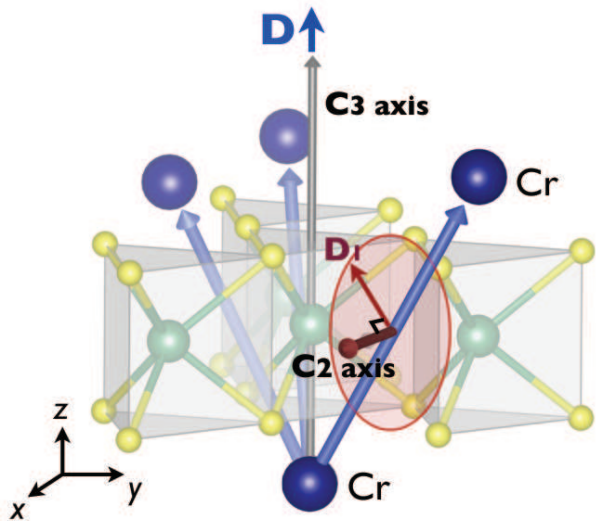}
\caption{(Color online) Illustration of the DM interaction in Cr$_{1/3}$NbS$_2$. Each Cr atom (blue spheres) is linked to three
Cr atoms in one of the adjacent planes (indicated by the blue arrows). The S and Nb atoms are the yellow and green spheres, respectivley.
The cylindrical red arrow represents a $C_2$ rotational symmetry axis located halfway between the two Cr atoms. The DM vector ${\bf D}_1$ 
linking these two Cr atoms (flat red arrow) lies in the red-shaded plane perpendicular to the $C_2$ axis. Note the precise orientation 
of ${\bf D}_1$ in the red-shaded plane is unknown. The related DM vectors
${\bf D}_2$ and ${\bf D}_3$ linking the lower Cr atom with the other two Cr atoms in the upper layer are not shown.
The gray arrow represents a $C_3$ rotational symmetry axis relating the three Cr atoms in the upper layer. The short blue arrow 
above this is the resultant DM vector ${\bf D} \! = \! {\bf D}_1 \! + \! {\bf D}_2 \! + \! {\bf D}_3$.}
\label{fig4}
\end{figure}

To provide a possible explanation for the enhanced mean magnetic field sensed by the muon ensemble ($\sim \! f$) 
at low temperatures, we adopt a Landau 
free energy analysis. In the absence of the ${\bf H} \! \perp \! \hat{c}$
magnetic field, the helimagnetic phase occurs via FM and DM interactions due to a lack of inversion symmetry. As shown in 
Fig.~\ref{fig4}, the Cr atoms form layered triangular lattices in Cr$_{1/3}$NbS$_2$, with an hexagonal space group of $P6_{3}22$. 
For each Cr atom in a layer, there are three nearest-neighbor Cr atoms in an adjacent layer. $C_2$ rotational symmetry exists
about an axis located halfway between two Cr atoms in the different layers, with inversion symmetry about this point broken by 
the S atoms. The DM vector linking the two Cr atoms (${\bf D}_1$) then lies in a plane perpendicular to 
the $C_2$ axis, as shown in Fig.~\ref{fig4}. Since the three Cr atoms within a layer are related by $C_3$ rotational 
symmetry (about an axis parallel to the $z$-axis), there are three DM vectors related by symmetry (${\bf D}_1$, ${\bf D}_2$, and ${\bf D}_3$),
which when summed together give a resultant DM vector ${\bf D}$ pointing along the $z$-axis. 
Consequently, the spin density of the helical order can be described by\cite{Bak:80,Plumer:81}
\begin{equation}
{\bf S}({\bf r}) = {\bf S} \exp(i{\bf Q} \cdot {\bf r}) + {\bf S}^* \exp(-i{\bf Q} \cdot {\bf r}) \, ,
\end{equation}
where ${\bf Q} \! = \! Q_0 \hat{z}$, ${\bf S} \! = \! S_x \hat{x} \! + \!  i S_y \hat{y}$, and by sysmmetry $|S_x| \! = \! |S_y| \! \equiv \! S$.
Since the free energy should be invariant under the exchange of ${\bf Q} \! \rightarrow \! -{\bf Q}$ and ${\bf S} \! \rightarrow \! -{\bf S}^*$,  
it is given to fourth order by\cite{Bak:80,Plumer:81}
\begin{equation}
F = \alpha_0 {\bf S} \cdot {\bf S}^* + DQ_0S^2 + \alpha_1 Q_0^2 S^2 + \beta S^4 \, ,
\end{equation}
where $D$ represents the strength of the DM interaction,\cite{Dzya:64} and the parameter $\beta$ is assumed to be positive, 
such that the free energy is bound. Minimizing $F$ with respect to $Q_0$, we find $Q_0 \! = \! -D/2\alpha_1$, resulting in the simplified free
energy equation
\begin{equation}
F = \alpha_Q S^2 + \beta S^4 \, ,
\end{equation}
where $\alpha_Q \! = \! \alpha_0 \! - \! D^2/4 \alpha_1$. Subsequent minimization of $F$ with respect to $S$, yields 
$S\! = \! \sqrt{- \alpha_Q/2 \beta}$. Hence the transition temperature $T_c(H \! = \! 0)$ below which helical magnetic
ordering occurs, is determined by when $\alpha_Q$ becomes negative.

A low-temperature enhancement of the local field sensed by the muon can be achieved via an increase of $S$, which implies
an increase in the negative value of $\alpha_Q$ and/or a decrease in the positive value of $\beta$. 
Since $\beta$ originates from the spin-spin interaction, smaller $\beta$ implies a stronger FM interaction between the Cr spins --- although 
this effect would be quite small. On the other hand, a larger negative value of $\alpha_Q$ corresponds to a weaker DM interaction $D$. 
This will occur if there is a change in the ratio between the $a$ (and $b$) and $c$-axis lattice constants, such that there is a rotation 
of the vectors ${\bf D}_1$, ${\bf D}_2$, and ${\bf D}_3$ towards the $xy$ plane. Thus we propose a change in the length of one or more
of the lattice constants below $T \! \sim \! 50$~K. 

In conclusion, our $\mu$SR measurements reveal a possible change in the magnetic structure of Cr$_{1/3}$NbS$_2$ below $T \! \sim \! 50$~K,
which is the same temperature region where anomalous behavior has been observed in bulk transport measurements.
The low-temperature modification of the magnetic structure can be explained by a change in lattice constants, which can be confirmed 
by precision measurements of the crystal structure. Moreover, to confirm that the change in lattice constants is such that it causes a reduction 
of the DM interaction, we suggest a neutron scattering measurement to look for a change of the ordering wavevector $Q_0$ --- 
since a consequence of a smaller DM interaction is a longer spin-helix period. Lastly, we do not find anything unusual about
the magnetism near the PM transition at low field (such as phase separation) that would contribute to the anomalies oberved 
in transport measurements near $T_c$.

\begin{acknowledgments}    
We thank the staff of TRIUMF's Centre for Molecular and Materials Science for technical assistance,
and Heungsik Kim for useful discussions. JES and HYK acknowledge support from CIFAR and NSERC of 
Canada. DGM and LL acknowledge support from the National Science Foundation (NSF-DMR-1410428). 
\end{acknowledgments}

\end{document}